\documentclass[12pt]{article}
\usepackage{amssymb}

\def\F{{\cal F}}

\newtheorem{theorem}{Theorem}
\newtheorem{prop}[theorem]{Proposition}

\begin{document}

\bigskip
\centerline{{\Large\bf Deformations of dispersionless}}

\bigskip

\centerline{{\Large\bf KdV hierarchies}}
\vspace{.3in}
\centerline{{\bf I.A.B. Strachan}}\vspace{.1in}
\centerline{Department of Mathematics, }
%\vspace{.1in}
\centerline{University of Hull,}
%\vspace{.1in}
\centerline{Hull, HU6 7RX,}
%\vspace{.1in}
\centerline{England.}
\vspace{.15in}
\centerline{e-mail: i.a.b.strachan@hull.ac.uk}

\vspace{.4in}
\centerline{{\bf Abstract}}

\vspace{.3in}
\small
\hskip 10mm\parbox{5.0in}{The obstructions to
the existence of a hierarchy of hydrodynamic conservation
laws are studied for a multicomponent dispersionless KdV system.
It is shown that if an underlying algebra is
Jordan, then the lowest obstruction
vanishes and that all higher obstructions automatically vanish.
Deformations of these multicomponent dispersionless
KdV-type equations are also studied. No new obstructions appear, and
hence the existence of a fully deformed hierarchy depends on the
existence of a single purely hydrodynamic conservation law.
}
\normalsize

\bigskip

\section{Introduction}

Consider the system of hydrodynamic type
\begin{equation}
u^i_t = a^i_{jk} u^j u^k_x\,,\quad\quad i\,,j\,,k=1\,,\ldots\,, N\,,
\label{dkdv}
\end{equation}
where the $a^i_{jk}$ are constant. These may be regarded as the structure
constants, with respect to some basis $e_i\,,$ of an algebra $\F\,,$
\[
e_i \circ e_j = a^k_{ij} e_k\,.
\]
Introducing an $\cal F$-valued field ${\cal U}=u^i e_i$ equation
(\ref{dkdv}) takes the simple form
\[
{\cal U}_t = {\cal U} \circ {\cal U}_x\,.
\]
In this paper the following conditions on the algebra will be assumed:
\begin{itemize}
\item[$\bullet$] $\F$ is a commutative algebra;
\item[$\bullet$] $\F$ possess a unity element $e_1\,,$ so $e_1\circ e_i=e_i\,;$
\item[$\bullet$] $\F$ is further equipped with a non-degenerate inner product
$<~,~>$
such that
\[
<a\circ b, c> = < a, b\circ c>
\]
This is known as the Frobenius condition.
\end{itemize}
In terms of local coordinates in which the metric is written as $\eta_{ij}\,,$
this last condition may be written as
\[
\eta_{in} a_{jk}^{n} = \eta_{kn} a_{ji}^{n}\,.
\]
which is equivalent to the condition that the tensor $a_{ijk}$ (where indices are
raised and lowered using $\eta$ and $\eta^{-1}$) is totally symmetric. Note that no
further conditions, such as the algebra being associative or Jordan, are imposed.
In what follows it will be useful to introduce the so-called
associator $\Delta_{ijk}^{~~~s}$ defined by
\[
(e_i \circ e_j) \circ e_k - e_i \circ (e_j \circ e_k)
=\Delta_{ijk}^{~~~s}e_s\,,
\]
or, in components, by
\[
\Delta_{ijk}^{~~~s} = a_{ij}^{~~r} a_{rk}^{~~s} - a_{jk}^{~~r}
a_{ir}^{~~s}\,.
\]
This is a measure of the deviation of the commutative algebra $\F$ from being associative.
Certain algebraic properties of the associator
will be required later, and these are given in the appendix.

With these conditions equation (\ref{dkdv}) may be written in Hamiltonian form
\[
u^i_t = \eta^{ij} \frac{d~}{dx} \frac{\delta~}{\delta u^j} \int h^{(3)} dx\,,
\]
where
\[
h^{(3)} = \frac{1}{3!} a_{ijk} u^i u^j u^k\,,
\]
or as $u^i_t = \{H^{(3)}, u^i\}$ where generically,
\[
H=\int h \,dx
\]
and the Hamiltonian structure is defined by
\[
\{H,G\} = \int \frac{\delta H}{\delta u^i}
\Big( \eta^{ij} \frac{d~}{dx} \Big)
\frac{\delta G}{\delta u^j} \,dx\,.
\]
The fact that this defines a Hamiltonian structure follows from the fundamental
theorem of Dubrovin and Novikov [DN].

\section{Generalized Symmetries and Jordan algebras}

Suppose one wishes to construct a hierarchy of hydrodynamic Hamiltonians which
commute with this lowest (non-trivial) Hamiltonian, that is, to find functionals
$H^{(n)}=\int h^{(n)}\,dx$ such that
\[
\{H^{(n)},H^{(3)}\} = 0\,.
\]
This implies that
\[
\int \frac{\partial h^{(n)}}{\partial u^i} a^i_{jk} u^h u^k_x\,dx = 0\,,
\]
so, under the assumption of rapidly decreasing, or periodic, boundary conditions, the
integrand must be a total $x$-derivative. Thus
\[
{\cal E} \Big\{\frac{\partial h^{(n)}}{\partial u^i} a^
i_{jk} u^h u^k_x\Big\} = 0
\]
where ${\cal E}$ is the Euler operator. This implies the following overdetermined
set of equations for the Hamiltonian density $h^{(n)}\,:$
\begin{equation}
a^i_{jk} u^j \frac{\partial^2 h^{(n)}}{\partial u^i \partial u^p} =
a^i_{jp} u^j \frac{\partial^2 h^{(n)}}{\partial u^i \partial u^k}\,.
\label{basicequation}
\end{equation}
It will be assumed that the densities are
homogeneous functions of degree $n\,,$ so by Euler's theorem
\[
E\left(h^{(n)}\right) = n h^{(n)}\,.
\]
where $E$ is the Euler vector field
\[
E=u^i \frac{\partial ~}{\partial u^i}
\]
As shown in [St], this, together with the earlier assumptions on the algebra
$\F$ imply the following recursion equations connecting the densities:
\begin{equation}
h^{(n+1)} = a^i_{jk} u^j u^k \frac{ \partial h^{(n)}}{\partial u^i}
\label{recursionup}
\end{equation}
and
\begin{equation}
\frac{\partial h^{(n)}}{\partial u^1} = h^{(n-1)}\,.
\label{recursiondown}
\end{equation}
The derivation of (\ref{recursionup}) (obtained by substituting
$k=1$ into (\ref{basicequation}) and using homogeneity) only uses a subset of the
equations in the overdetermined systems (\ref{basicequation}). One
must therefore check that the $h^{(n)}$ obtained by the recursive
application
of (\ref{recursionup}) does satisfy the full overdetermined system.
In general there is an obstruction:

\begin{prop} Suppose that $h^{(n)}$ is a conserved density. The
the function $h^{(n+1)}$ obtained by using (\ref{recursionup}) is
a conserved density if and only if
\begin{equation}
\Delta_{irj}^{~~~s} u^r \frac{\partial h^{(n)}}{\partial u^s} =
0\,.
\label{obstruction}
\end{equation}
Thus for an associative algebra all obstructions vanish.
\end{prop}

\medskip

\noindent Equation (\ref{obstruction}) may be viewed as extra
generalized homogeneity conditions on the densities. Alternatively
they may be viewed as forcing the flows to live in some reduced
ring. The following results is easily obtained by direct
calculation. Let
\[
\Xi_{ij}=\Delta_{irj}^{~~~s} u^r \frac{\partial ~}{\partial
u^s}\,.
\]
Then:
\begin{eqnarray*}
\Xi_{1i} & = & 0\,,\\
\Xi_{ij} + \Xi_{ji} & = & 0\,,\\
{[}\Xi_{ij},E{]} & = & 0 \,,\\
{[} \Xi_{ij},\Xi_{rs} {]} & = &
( \Delta_{ipj}^{~~~q} \Delta_{rqs}^{~~~n}-\Delta_{rps}^{~~~q}
\Delta_{iqj}^{~~~n})
u^p \partial_n\,.
\end{eqnarray*}
For an arbitrary algebra $\F$ equation (\ref{obstruction}) will not have a
solution. However, a necessary condition for the existence of
a solution to (\ref{obstruction}), and hence for an
unobstructed hierarchy is given by Frobenius' Theorem. Thus
a necessary condition for the existence of
a solution to (\ref{obstruction}) is that the commutator of the
vector fields $\Xi_{ij}$ must be a linear combination of the
same vector fields:
\begin{equation}
{[} \Xi_{ij}, \Xi_{rs} {]} = c^{pq}_{ij,rs} \,\Xi_{pq}\,.
\label{commutator}
\end{equation}

\medskip

\noindent{\bf Example} Let $\F$ be the Jordan algebra $D_N$ defined by
\begin{eqnarray*}
e_1 \circ e_i & = & + e_i \,, \\
e_i \circ e_i & = & - e_1 \,, \quad i=2\,,\ldots\,, N\,, \\
e_i \circ e_j & = & 0 \,,\quad {\rm otherwise}\,.
\end{eqnarray*}
(Such a multiplication comes from an underlying Clifford
multiplication). Then
\[
\Xi_{ij} = \cases{ 0 & if $i$ or $j=1$ \,, \cr
u^i \partial_j - u^j \partial_i & otherwise}
\]
then, for $i\,,j\,,r\,,s\neq 1\,,$
\[
{[} \Xi_{ij}, \Xi_{rs} {]} = \delta_{jr} \Xi_{is} + \delta_{is}
\Xi_{jr} - \delta_{sj} \Xi_{ir} - \delta_{ri} \Xi_{js}\,,
\]
and so the necessary condition is satisfied. Note that the vector fields
define a Lie algebra isomorphic is $u(1) \oplus so(N-1)\,.$ In the case the
condition is also sufficient [St], for example when $N=3$ they
imply that
\[
h^{(n)} = h^{(n)} [\, u^1 \,, (u^2)^2+(u^3)^2 \,]\,.
\]
\bigskip

\noindent{\bf Example} The condition (\ref{commutator}) is not
sufficient to ensure a solution to equation (\ref{basicequation}).
Starting with the cubic prepotential
\[
F=\frac{1}{6} u^3 + \frac{1}{2} u (v^2+w^2) +
\frac{1}{6} \,\alpha_1\, v^3 + \frac{1}{2}\,\alpha_2\, v^2 w +
\frac{1}{2}\,\alpha_3\, v w^2 + \frac{1}{6}\,\alpha_4\, w^3
\]
one may define the algebra $\F$ by the equations
\[
a_{ijk} = \frac{\partial^3 F}{\partial u^i \partial u^j \partial
u^k}\,,
\]
with $\eta_{ij}=a_{1ij}\,.$ This, together with its inverse, is
then used to lower and raise indices, so $a_{jk}^i = \eta^{ir}
a_{rjk}\,.$ All algebras $\F$ arise in this way [St].
The only non-zero vector fields are
\[
\Xi_{23}=-\Xi_{32} = \Delta (v \partial_w-w\partial_v)\,,
\]
where $\Delta=\alpha_2^2 - \alpha_1 \alpha_3 + \alpha_3^2 - \alpha_2
\alpha_4-1\,,$ and hence the conditions in Proposition 2 are
trivially satisfied.
Since $h^{(2)}$ satisfies $\Xi_{23} (h^{(2)})=0$
then $h^{(3)}$ is a conserved density (as was known already).
However, $\Xi_{23}(h^{(3)})=0$ if and only if:

\medskip

(a) $\Delta=0\,;$

\medskip

\noindent or

(b) $\alpha_1=\alpha_2=\alpha_3=\alpha_4 = 0\,.$

\medskip

\noindent Algebraically these mean the algebra $\F$ is either

\medskip

(a) associative (and hence trivially Jordan)

\medskip

\noindent or

\medskip

(b) the Jordan algebra $D_3\,.$

\medskip

\noindent Clearly if $\F$ is associative then the
$h^{(n)}$ constructed recursively via (\ref{recursionup})
are all conserved densities. If $\F$ is the Jordan
algebra $D_3$ then
\[
\Xi_{23}\, h^{(n+1)} = 2 u \Xi_{23}\, h^{(n)}
\]
and hence all the obstructions vanish.

\medskip

The results obtained in the above example hold more generally.
Consider the obstruction to the existence of $h^{(2)}\,:$

\begin{eqnarray*}
\Xi_{ij}\, h^{(1)} & = & \Delta_{irj}^{~~~s} u^r
\partial_s ( \eta_{1p} u^p) \,, \\
& = & \Delta_{irj1} u^r \,, \\
& = & 0 \,.
\end{eqnarray*}
Hence $h^{(2)}$ exists (which was already known). Note
that the Casimirs $h^{(1)}_{\bullet}=\eta_{\bullet p} u^p\,,$ where $\bullet=1\,,\ldots \,,
N$ are all conserved densities. In the case when $\F$ is
associative, when all obstructions vanish, these given rise, via
(\ref{recursionup}), to $N$-independent families of conserved densities
$h^{(n)}_{\bullet}\,.$ For an arbitrary algebra it is only when $\bullet=1$ that
an $h^{(2)}$ exists automatically. The obstruction to the existence
of $h^{(3)}$ is similar:
\begin{eqnarray*}
\Xi_{ij}\, h^{(2)} & =& \Delta_{irj}^{~~~s} u^r \partial_s
(\frac{1}{2} \eta_{pq} u^p u^q ) \,, \\
 & = & \Delta_{ipjq} u^p u^q \,, \\
 & = & 0
\end{eqnarray*}
by symmetry/antisymmetry. Hence $h^{(3)}$ exists (which again was already known).
However there is an obstruction to the existence of $h^{(4)}\,:$
\[
\Xi_{ij} \, h^{(3)} = \frac{1}{2} c_{pq}^{~~s}
\Delta_{ris}^{~~~j} u^r u^p u^q
\,.
\]
Hence
\[
\left\{ \Xi_{ij}\, h^{(3)} \right\} \Longleftrightarrow
\left\{ c_{(pq}^{~~s} \Delta_{r)is}^{~~~j} = 0\right\}\,.
\]
However this is just the condition for the algebra $\F$ to be
Jordan. Hence:
\begin{prop}
\[
\left\{ ~h^{(4)}~{\rm exists}~\right\} \Longleftrightarrow
\left\{~\F~{\rm is~Jordan}~\right\}\,.
\]
\end{prop}

\bigskip

\noindent Note, it is not immediately clear that, even in
$h^{(4)}$ exists, $h^{(n)}$ exists for arbitrary $n\,;$
higher-order conditions on the algebra ${\cal F}$ could arise.
However this is not the case.

\begin{prop}
\[
\left\{~\F~{\rm is~Jordan}~\right\} \Longleftrightarrow
\left\{ ~h^{(n)}~{\rm exists~ for~}\forall n~\right\} \,.
\]
\end{prop}

\medskip

\noindent{\bf Proof} It will be useful to define
\[
\partial {\bf h}^{(n)} = \eta^{rs} \partial_r h^{(n)} e_s\,,
\]
so, from equation \ref{recursionup}
\[
(n-1) \partial {\bf h}^{(n)} = {\cal U} \circ \partial {\bf h}^{(n-1)}
\]
and hence by induction
\begin{equation}
\partial {\bf h}^{(n+1)} = \frac{1}{n!} {\cal U}^n
\label{dhn}
\end{equation}
where ${\cal U}^n = {\cal U} \circ {\cal U}^{n-1}$ (at this stage
there is no assumption of power associativity on ${\cal F}$). With
this the
obstruction (\ref{obstruction}) may be written as an algebraic
condition on the algebra ${\cal F}\,.$ Let ${\cal V}=v^i
e_i\,,{\cal W}=w^ie_i\,,$ and consider
\[
{\cal O}^{(n+1)} = v^i w^j \Delta_{irj}^{~~~s} u^r \frac{\partial
h^{(n)}}{\partial u^s}
\]
(i.e. ${\cal O}^{(n+1)}$ is the obstruction to the existence of
$h^{(n+1)}$). Using this definition and (\ref{dhn})
\begin{eqnarray*}
{\cal O}^{(n+1)} & = & \frac{1}{(n-1)!}
\langle
({\cal V}\circ {\cal U}) \circ {\cal W} -
{\cal V}\circ ({\cal U} \circ {\cal W}),
{\cal U}^{n-1}\rangle\,, \\
& = & \frac{1}{(n-1)!} \langle {\cal W}, ({\cal V}\circ {\cal
U})\circ{\cal U}^{n-1} - ({\cal V}\circ {\cal U}^{n-1})\circ {\cal
U}\rangle\,.
\end{eqnarray*}
on using the Frobenius condition. Since ${\cal W}$ is arbitrary,
it follows from the non-degeneracy of the inner product that
\[
\left\{~{\cal O}^{(n+1)}=0~\right\} \Longleftrightarrow
\left\{ ~
({\cal V}\circ {\cal U})\circ{\cal U}^{n-1} =
({\cal V}\circ {\cal U}^{n-1})\circ{\cal U}
~\right\} \,.
\]
If $n=3$ one just recovers proposition 3. However, if ${\cal F}$
is a
Jordan algebra then
\[
({\cal V}\circ {\cal U})\circ{\cal U}^{n-1} =
({\cal V}\circ {\cal U}^{n-1})\circ{\cal U} \quad \forall n \geq 2
\]
automatically. The proof of this fact is not immediately obvious
(except
in the case of special Jordan algebras) [Sc]. Hence the result.

\bigskip

Thus one has, in general, an infinite number of obstructions
${\cal O}^{(n)}$ to the existence of $h^{(n)}\,,$ but if
${\cal O}^{(4)}=0$ then all higher obstructions vanish
automatically.
So, if and only if $\cal F$ is a Jordan algebra, starting from
$h^{(3)}$ (or even from $h^{(2)}$ which generates the trivial flow)
one may construct recursively all the Hamiltonians starting from
the Casimir $h^{(1)}$

\[
h^{(1)}\rightleftarrows
\cdots \rightleftarrows h^{(n)} \begin{array}{c}
{}_{(\ref{recursionup})}
\\ \rightleftarrows \\ {}^{(\ref{recursiondown})} \end{array} h^{(n+1)}
\rightleftarrows \cdots\,.
\]

\noindent Having established conditions for the integrability of equation (\ref{dkdv})
integrable deformations of this hydrodynamic system will now be considered.
It will turn out the no new algebraic conditions will be required.

\section{Deformations of hydrodynamic flows}

\subsection{First Order Deformations}

Consider the dispersive KdV system
\begin{equation}
u^i_t = a^i_{jk} u^j u^k_x + \varepsilon u^i_{xxx}
\label{kdv}
\end{equation}
where $\varepsilon$ is a formal parameter. This too may be written
in Hamiltonian form
\[
u^i_t = \eta^{ij} \frac{d~}{dx} \frac{\delta~}{\delta u^j} \int
\Bigg\{ h^{(3)} - \varepsilon\Big(\frac{1}{2} \eta_{ij} u^i_x u^j_x \Big)\Bigg\}\,dx\,.
\]
Thus the dispersionless Hamiltonian density undergoes a first order deformation
\[
h^{(3)} \mapsto h^{(3)}(\varepsilon) = h^{(3)} -\varepsilon\Big(\frac{1}{2} \eta_{ij} u^i_x u^j_x \Big)\,.
\]
The higher Hamiltonian densities, if they are to commute with the
above deformed Hamiltonian will also undergo such a deformation,
though this will contain higher-order terms, not just first-order
terms,
\[
h^{(n)} \mapsto h^{(n)}(\varepsilon) = h^{(n)} +
\sum_{m=1}^\infty \varepsilon^m \delta^m h^{(n)}\,.
\]
The main result of this section is the following recursion formula for $\delta h^{(n)}\,:$
\begin{equation}
n\chi^{(n)}_{ij} = a_{ir}^{~~s} u^r \chi^{(n-1)}_{sj} - \frac{3}{2}
\frac{\partial^2 h^{(n-1)}}{\partial u^i \partial u^j}\,,
\label{firstorderrecursion}
\end{equation}
where $\delta h^{(n)} = \chi^{(n)}_{ij} u^i_x u^j_x\,.$ The motivation for these
calculations, and
comments on the form of higher order deformations, will be postponed until later.

To second order
\[
\{ H^{(n)} + \varepsilon \delta H^{(n)} ,  H^{(3)} + \varepsilon \delta H^{(3)} \} =
O(\varepsilon^2)\,.
\]
The zeroth order terms have already been constructed and the first order deformation
$\delta H^{(n)}$ must satisfy
the equation
\[
\{  H^{(n)},  \delta H^{(3)}\} +
\{  \delta H^{(n)} , H^{(3)}\} = 0 \,.
\]
Unpacking the various definition yields, on integrating by parts once to eliminate
terms involving $u^i_{xxx}$, again using rapidly decreasing, or periodic, boundary conditions:
\begin{equation}
\int
\Big\{A^{(n)}_{ij} u^i_{xx} u^j_x + B^{(n)}_{ijk}u^i_x u^j_x u^k_x \Big\}
\, dx = 0
\label{integrand}
\end{equation}
where
\begin{equation}
A^{(n)}_{ij}=-\partial_i\partial_j h^{(n)} -2 a^r_{sj} u^s \chi^{(n)}_{ir} \,, \quad\quad
B^{(n)}_{ijk}  = \frac{1}{3!} \sum_{(ijk)\in S_3} a^r_{sk} u^s
\Big\{ \frac{\partial \chi^{(n)}_{ij}}{\partial u^r} - 2
\frac{\partial \chi^{(n)}_{ri}}{\partial u^j}\Big\}\,.
\label{defaandb}
\end{equation}
It follows from the homogeneity of the densities $h^{(n)}$
that the $\chi^{(n)}_{ij}$ are homogeneous of degree $n-3\,,$ so
\begin{equation}
u^r \frac{\partial\chi^{(n)}_{ij}}{\partial u^r} = (n-3) \chi^{(n)}_{ij}\,,
\label{firstorderhomog}
\end{equation}
and from the existence of the unity element in the algebra $\F$ it follows
that
\begin{equation}
\frac{\partial\chi^{(n)}_{ij}}{\partial u^1}=\chi^{(n-1)}_{ij}\,.
\label{firstorderdown}
\end{equation}
For the integrand in (\ref{integrand}) to be a total
$x$-derivative requires the following conditions\footnote{Equation
(\ref{a}) has been used in the derivation of (\ref{b}).}, obtained
using the Euler operator ${\cal E}\,:$
\begin{eqnarray}
A^{(n)}_{ij} & = & A^{(n)}_{ji} \,, \label{a}\\
6 B^{(n)}_{ijk} & = & A^{(n)}_{ij,k} + A^{(n)}_{jk,i} + A^{(n)}_{ki,j}\label{b}
\end{eqnarray}
The procedure to solve this overdetermined system is to
use equations (\ref{defaandb}-\ref{b}) with special values of
$i\,,j\,,k$ to derive (\ref{firstorderrecursion}), and then find
conditions, if any, on the constants $a^i_{jk}$ so that it solves
the full set of equations.

Equation (\ref{b}) simplifies to
\[
\sum_{{\rm cyclic}~i\,,j\,,k}
a^r_{sk} u^s \Big\{
\partial_r \chi^{(n)}_{ij} - \partial_j \chi^{(n)}_{ri}
\Big\} + a^r_{ij} \chi^{(n)}_{kr} = -\frac{3}{2} \partial_i \partial_j \partial_k h^{(n)}\,.
\]
If $i=j=k=1$ one obtains
\[
u^r \{\partial_r \chi^{(n)}_{11} - \partial_1 \chi^{(n)}_{1r} \} + \chi_{11}^{(n)}
= -\frac{1}{2} h^{(n-3)}\,.
\]
If $i=j=1\,,k\neq 1$ one obtains, on using the relation
$A_{1k}=A_{k1}\,,$
\[
2(n-1) \chi_{1k}^{(n)} - \partial_k ( u^r \chi_{r1}^{(n)}) =
-\frac{3}{2} \partial_k h^{(n-2)} + 2 a^r_{sk} u^s
\chi_{r1}^{(n-1)} - a^r_{sk} \partial_r \chi_{11}^{(n)}\,.
\]
Assuming that
\[
\chi_{11}^{(n)} = -\frac{1}{2} h^{(n-3)}
\]
one finds that
\[
u^r \chi_{r1}^{(n)}= -\frac{1}{2} (n-2) h^{(n-2)}
\]
and hence the recursion relation
\[
(n-1) \chi_{k1}^{(n)} = - \partial_k h^{(n-2)} + a^r_{sk} u^s
\chi_{r1}^{(n-1)}\,.
\]
Using the initial condition $\chi_{ij}^{(2)}=-1/2\, \eta_{ij}$ one
may then
prove by induction that
\[
\chi_{k1}^{(n)} = -\frac{1}{2} \partial_k h^{(n-2)}\,.
\]
Finally, if $i=1\,,j\,,k\neq 1\,,$ one obtains the required relation
(\ref{firstorderrecursion}) on using the derivative of the relation
$A_{1k}=A_{k1}\,.$
The first few solutions, starting with are
$\chi^{(2)}_{ij}=0\,,$ are
\[
\chi^{(3)}_{ij}=-\frac{1}{2} \eta_{ij}\,,\quad\quad
\chi^{(4)}_{ij}=-\frac{1}{2} c_{ijr} u^r\,, \quad\quad
\chi^{(5)}_{ij}=-\frac{1}{20} \big\{ a^q_{ij} a_{qrs} + 4 a^q_{ir}
a_{qjs} \big\} u^r u^s\,.
\]
These results, and more generally the recursion formula
(\ref{firstorderrecursion}) have been derived using a subset of
the full governing equations, and so one must check whether or
not they satisfy the full set of equations. This will be
done below.

\subsection{Higher Order Deformations}

In principle one may continue this procedure to find, if they
exist, second order and higher deformations $\delta^i H^{(n)}$ to
the unperturbed Hamiltonians. These would have to satisfy the
equation
\begin{equation}
\{\delta^i H^{(n)},H^{(3)}\}+\{\delta^{i-1} H^{(n)},\delta
H^{(3)}\} = 0\,.
\label{higherorder}
\end{equation}
The number of terms in such deformations grows very rapidly. For
$i=2\,,$
\begin{equation}
\delta^{(2)} H^{(n)} = \Xi_{ij}^{(n,0)} u^i_{xxx} u^j_x +
\Xi^{(n,1)}_{ij} u^i_{xx} u^j_{xx} + \Xi^{(n,2)}_{ijkl}u^i_x u^j_x
u^k_x u^k_x\,,
\label{deltasquared}
\end{equation}
and in general the number of such terms grows as the number of
partitions of $n\,.$ For $n=4$ and $5$ equation
(\ref{higherorder}) may be solved perturbatively with no
constraint
on the algebra $\F\,,$ the results being
\[
h^{(4)}(\varepsilon) = h^{(4)} +
\Big( -\frac{1}{2} a_{ijk} u^i u^j_x u^k_x\Big) \varepsilon+
\Big(\frac{3}{10} \eta_{ij} u^i_{xx} u^j_{xx} \Big) \varepsilon^2
\]
and $h^{(5)}(\varepsilon)=$
\[
h^{(5)} +
\Big(-\frac{1}{20} \big\{ a^q_{ij} a_{qrs} + 4 a^q_{ir}
a_{qjs} \big\} u^r u^s u^i_x u^j_x\Big) \varepsilon +\\
\Big( \frac{3}{10} a_{ijk} u^i u^j_{xx} u^k_{xx} \Big)
\varepsilon^2 +
\Big( -\frac{9}{70} \eta_{ij} u^i_{xxx} u^j_{xxx} \Big)
\varepsilon^3\,.
\]

\subsection{Vanishing Obstructions and the Recursion Operator}

Proceeding in this way one would expect new obstructions appearing
at each order. However, if ${\cal F}$ is Jordan then there are no
new obstructions at any order. This follows, not directly, as in
section 2, but from the use of a recursion operator. The algebra
$\cal F$ may be used to define an operator
\[
{\cal R}^i_j = \delta^i_j \Bigg( \frac{d~}{dx}\Bigg)^2 +
\Bigg\{
    \frac{2}{3} a_{jk}^{~~i} u^k + \frac{1}{3} a_{jk}^{~~i} u_x^k
    \Bigg(\frac{d~}{dx}\Bigg)^{-1}
\Bigg\} +
\frac{1}{9}  \Delta_{kl}^{~~ji}
u^l \Bigg( \frac{d~}{dx} \Bigg)^{-1} \Bigg\{ u^k \Bigg(
\frac{d~}{dx}\Bigg)^{-1}\Bigg\}\,,
\]
and it has been shown in [GK,Sv] that this defines a recursion
operator if and only if ${\cal F}$ is Jordan. Hence one may obtain
a bi-Hamiltonian hierarchy
\[
{\bf u}_{t_n} = {\cal R}^n {\bf u}_x\,.
\]
Note however, that the condition that ${\cal F}$ is Jordan comes
from the existence of a hydrodynamic
conservation at the lowest possible degree at zeroth order; that
is, it is the underlying dispersionless equation (\ref{dkdv})
which
governs the properties and existence of the full dispersive
hierarchy.
Thus
\[
\left\{
\matrix{ {\rm The~existence~of~the~purely}\cr
{\rm hydrodynamic~density~}h^{(4)}\cr}
\right\}
\,\Longleftrightarrow
\left\{
\matrix{ {\rm The~existence~of~a~fully}\cr
{\rm deformed~dispersive~hierarchy}\cr
{\rm at~all~order}\cr}
\right\}\,,
\]
there being no new obstruction.

\section{Comments}

The motivation of this paper came from the work of Eguchi et. al.
[EYY] who studied deformations of hydrodynamic systems associated
to topological field theories, via Frobenius manifolds. They
found, in specific examples, that first order deformations always
exist, but obstructions will in general occur at second order.
This has now been proved in general for semi-simple Frobenius
manifolds [DZ]. Little work has been done on the existence of
higher-order deformations for systems other than the KdV equation
[DM]. Artificial examples may be constructed by scaling known
dispersive hierarchies, but this does not resolve the basic
problem of when a given hydrodynamic system may be deformed, or
conversely, is the dispersionless limit of an integrable
dispersive hierarchy.

The multicomponent KdV-type equations in this paper are not
associated to Frobenius manifolds, unless the algebra $\F$ is
associative [St], though they may be formulated in terms of a Jordan manifold [St].
However the same generic features remain. The
equations for first order deformations, although overdetermined,
still admit a solution. Pivotal in this derivation is the existence of a unity
element in the algebra $\F\,;$ for Frobenius manifolds this is
automatic.

The form of these results bear a close resemblance to ideas used
in deformation quantization; first order deformations define a
$\partial$-operator, and the obstructions at higher order are
measured by a cohomology group. One might therefore expect that
the results of this paper, and those of [EYY,DM,DZ], could be
given some cohomological interpretation with the Jordan condition being equivalent to
the vanishing of some cohomology group. Alternatively, the
overdetermined systems (\ref{basicequation}) could be written as
an exterior differential system and its integrability expressed in
terms of vanishing curvature and torsions.

\section*{Appendix}

The following results follow immediately from the definition of
the associator:

\begin{eqnarray*}
\Delta_{ijk}^{~~~s} + \Delta_{kji}^{~~~s} & = & 0 \,, \\
\Delta_{ijk}^{~~~s} + \Delta_{jki}^{~~~s} + \Delta_{kij}^{~~~s} &
= & 0 \,.
\end{eqnarray*}
The tensor $\Delta_{ijks}$ is defined by
\begin{eqnarray*}
\Delta_{ijks} & = & \langle \Delta_{ijk}^{~~~r} e_r, e_s\rangle\,,
\\
& = & \Delta_{ijk}^{~~~r} \eta_{rs} \,.
\end{eqnarray*}
Using the Frobenius conditions
\begin{eqnarray*}
\Delta_{ijks} & = &
\langle (e_i \circ e_j) \circ e_k,e_s\rangle -
\langle e_i \circ (e_j \circ e_k),e_s\rangle\,, \\
& = & \langle e_i \circ e_j, e_k \circ e_s \rangle -
\langle e_j \circ e_k , e_i \circ e_s \rangle\,.
\end{eqnarray*}
From this and the commutativity of the algebra the following to results are immediate:
\begin{eqnarray*}
\Delta_{ijks} & = & \phantom{-} \Delta_{jisk} \,, \\
\Delta_{ijks} & = & - \Delta_{iskj}\,.
\end{eqnarray*}
These results show that the associator is an algebraic curvature
tensor under the identification
\[
R^s_{~jik} \longleftrightarrow \Delta_{ijk}^{~~~s}\,.
\]
More specifically, the pair $\{ \eta_{ij}, c_{ij}^k \}$ define
a metric and a torsion free (though not metric) connection. The
Bianchi identify will automatically be satisfied, but this may also be
checked by direct calculation.

One further results will be required:
\begin{eqnarray*}
\Delta_{ijk1} & = & c_{ij}^{~~r} c_{rk1} - c_{jk}^{~~r} c_{ri1}
\,, \\
& = & c_{ij}^{~~r} \eta_{rk} - c_{jk}^{~~r} \eta_{ri} \,, \\
& = & c_{ijk} - c_{jki} = 0 \,.
\end{eqnarray*}

\section*{Bibliography}

\begin{itemize}

\item[{[DM]}] {Dubrovin, B. and Maltsev, A. Ya., {\sl Recurrent
procedure for the determination of the Free Energy
$\epsilon^2$-expansion in the Topological Field Theory,} solv-int/9904004.}

\item[{[DN]}] {Dubrovin, B. and Novikov, S.P., {\sl The Hamiltonian formalism of one-dimensional
systems of hydrodynamic type and the Bogolyabov-Witham averaging method},
Soviet Math. Dokl. {\bf 27} (1983) 665-669.}

\item[{[DZ]}] {Dubrovin, B. and Zhang, Y.,
{\sl Bihamiltonian hierarchies in the 2D Topological
Field Theory at One-Loop Approximation}, C.M.P. {\bf 198} (1998) 311-361,
{\sl Frobenius Manifolds and Virasoro Constraints}, math/9808048.}

\item[{[EYY]}] {Eguchi, T, Yamada Y. and Yang, S.K., {\sl On the genus expansion in the
topological string theory}, Rev. Math. Phys. {\bf 7} (1995) 279-309.}

\item[{[GK]}] {G\"urses, M. and Karasu, A., {\sl Integrable coupled KdV systems},
Phys. Lett. {\bf A 39} (1998) 2103-2111.}

\item[{[Sc]}] {Schafer, R.D., {\sl An Introduction to Nonassociative Algebras},
(Academic Press, New York, 1966).}

\item[{[St]}] {Strachan, I.A.B., {\sl Jordan manifolds and
dispersionless KdV equations}, to appear in {\sl Journal of Nonlinear Mathematical Physics}.}

\item[{[Sv]}] {Svinolupov, S.I., {\sl Jordan algebras and generalized Korteweg-de Vries
equations}, Teor. Math. Phys. {\bf 87} (1991) 611-620, {\sl Jordan algebras and integrable
systems}, Funct. Anal. and its Appl. {\bf 27} (1993) 257-265.}

\end{itemize}

\end{document}